\documentclass[a4paper]{jpconf}

\usepackage{graphicx}
\usepackage{color}

\usepackage{dcolumn}
\usepackage{amssymb}

\begin{document}
\title{Fitting BICEP2 with defects, primordial gravitational waves and dust}
\newcommand{\addressSussex}{Department of Physics \&
Astronomy, University of Sussex, Brighton, BN1 9QH, United Kingdom}
\newcommand{\addressGeneva}{D\'epartement de Physique Th\'eorique \& Center for Astroparticle Physics,
Universit\'e de Gen\`eve, Quai E.\ Ansermet 24, CH-1211 Gen\`eve 4, Switzerland}
\newcommand{\addressBilbao}{Department of Theoretical Physics, University of the Basque Country UPV/EHU,
48080 Bilbao, Spain}
\newcommand{\addressEdinburgh}{Institute for Astronomy, University of Edinburgh, Royal Observatory, Edinburgh EH9 3HJ,
United Kingdom}
\newcommand{\addressHelsinki}
{Department of Physics and Helsinki Institute of Physics, PL 64, FI-00014 University of Helsinki, Finland}
\newcommand{\addressAIMS}
{African Institute for Mathematical Sciences, 6 Melrose Road, Muizenberg, 7945, South Africa}

\author{
Joanes Lizarraga$^{1}$,
Jon Urrestilla$^1$,
David Daverio$^{2}$,
Mark Hindmarsh$^{3,4}$, 
Martin Kunz$^{2,5}$,
Andrew R.~Liddle$^{6}$}

\address{
$^1$ \addressBilbao \\
$^2$ \addressGeneva \\
$^3$ \addressSussex \\
$^4$ \addressHelsinki \\
$^5$ \addressAIMS \\
$^6$ \addressEdinburgh}
\ead{joanes.lizarraga@ehu.es}

\begin{abstract}
In this work we discuss the possibility of cosmic defects being responsible for the B-mode signal measured by the BICEP2 collaboration. We also allow for the presence of other cosmological sources of B-modes such as inflationary gravitational waves and polarized dust foregrounds, which might contribute to or dominate the signal. On the one hand, we find that defects alone give a poor fit to the data points. On the other, we find that defects help to improve the fit at higher multipoles when they are considered alongside inflationary gravitational waves or polarized dust. Finally, we derive new defect constraints from models combining defects and dust. This proceeding is based on previous works \cite{Lizarraga:2014eaa,Lizarraga:2014xza}. 
\end{abstract}

\section{Introduction}

The recent detection of B-mode polarization on large angular scales \cite{Ade:2014xna} has opened a new window to test and constrain models that predict primordial perturbations. The leading candidate, as claimed by the BICEP2 team, is primordial inflationary gravitational waves. For a tensor-to-scalar ratio $r$ of around 0.2, these give a good match to the spectral shape in the region $\ell \approx [40\ 150]$.

An alternative mechanism of generating primordial B-modes is the presence of cosmic defects. Even though their relative contribution to the temperature power spectrum is expected to be sub-dominant, they can still contribute importantly to the B-mode polarization. We explore whether cosmic defects could explain or help other primary contributors fit the data better.

Recently some works have reported that the measurements made by BICEP2 can have a non-negligible astrophysical source: polarized dust foregrounds \cite{Mortonson:2014bja,Flauger:2014qra,Adam:2014bub}. We extended our primary analysis including such a possible source.

\section{Cosmic defects and defect zoo}

Cosmic defects are extended objects that could be produced during cosmological phase transitions at the earliest stages of our universe, when spontaneous symmetry breaking occurred. Defects appear as a consequence of the development of a non trivial vacuum configuration.

Depending on the symmetry of the system, as well as on the vacuum manifold configuration, the breaking process leads to different defect networks. Among the wide range of defects analyzed in the literature, we have focused on the following types, which can be considered as the representative ones:

\begin{enumerate}
\item Abelian Higgs strings (AH strings): consequence of a broken $U(1)_L$ local symmetry.
\item Textures (TX): products of the breaking of a global $O(4)_G$ symmetry.
\item Semi-Local strings (SL strings): produced after a $SU(2)_G\times U(1)_L$ symmetry is broken. In some sense, they are a mixture of local and global defects.
\end{enumerate}

Cosmic defects are predicted by many inflationary scenarios and their cosmological implications have been widely analyzed. Cosmic defects not only generate CMB anisotropies, but they are also candidates for the generation of other phenomena such as gravitational waves or lensing. Their detection, therefore,  would give invaluable information about the physics of the early universe.

\subsection{CMB anisotropies}

The possible contribution of defects to CMB anisotropies has been one of their most studied observable imprints, both in temperature and polarization channels. Their contribution has been highly constrained via the temperature anisotropies \cite{Urrestilla:2011gr,Ade:2013zuv,Lizarraga:2012mq}.

In contrast to ordinary inflationary perturbations, which are set by the initial perturbations, scaling of defects implies that they continuously induce perturbations to the background metric, \textit{i.e.} they are active sources of perturbations.

One of the most important differences between primordial and defect induced anisotropies arises in the vector perturbations. Inflationary vector modes decay due to the cosmological expansion and do not create B-modes. On the contrary, defect vector perturbations are seeded continuously and consequently do not decay. The contribution of defects to B-modes, therefore, is twofold: through tensor \textit{and} vector perturbations; and both ingredients contribute in a comparable amount. Scalar perturbations, in turn, do not induce directly B-modes, they only contribute through lensing of E-modes. This is the reason why it is plausible that the relative contribution of inflation and defects to B-modes can be similar in amplitude.


All defect networks produce CMB anisotropies in a very similar way, and even though the shape of the power spectra is very similar, there are some differences that must be pointed out. Beyond the shape and peak position, those differences come from the constraints imposed by CMB experiments. In opposition to the inflationary case, where scalars and tensors can vary almost independently (apart from the inflationary consistency relation), the amplitudes of temperature and polarization power spectra from defects are strictly related. Thus, though no experiment had constrained defects in the BB channel until BICEP2 released their data, the possible defect contribution to it has already been limited by the temperature constraints.

\begin{figure}[h!]
{\includegraphics[width=0.5\textwidth]{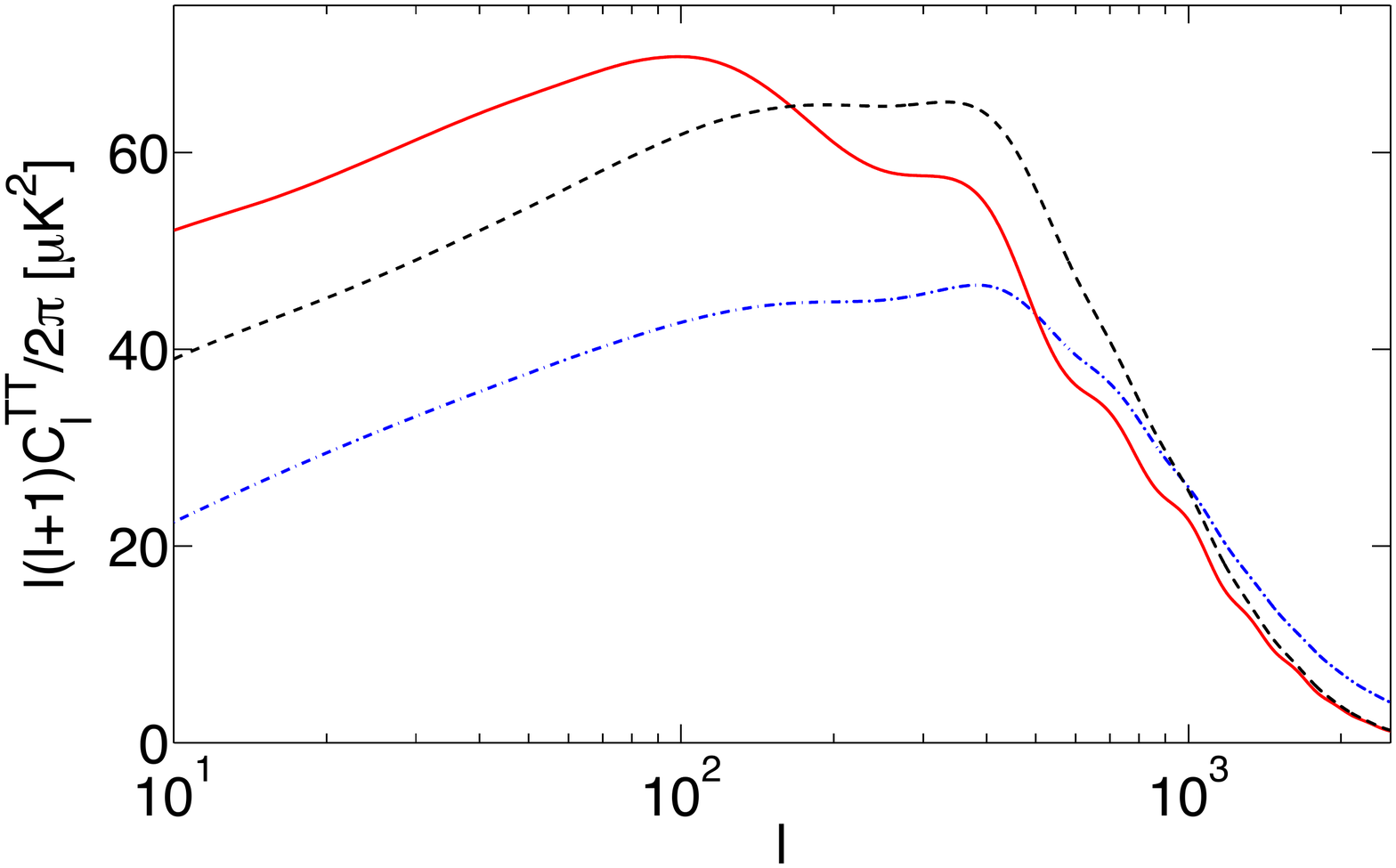}}
{\includegraphics[width=0.5\textwidth]{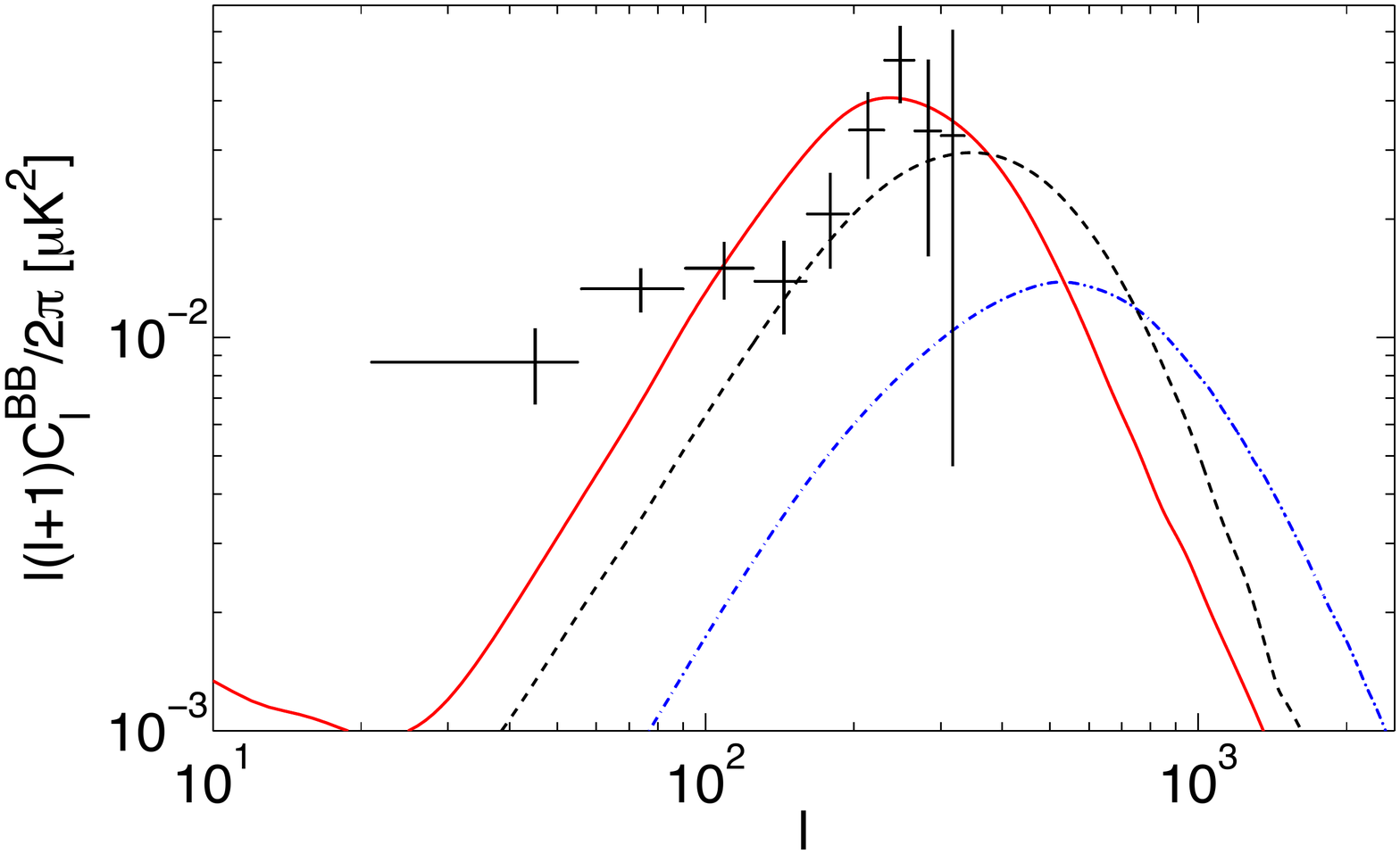}}
\caption{\label{95_Defects} Defect spectra normalised to the $95\%$ upper limits obtained using {\it Planck} + WMAP (EE and TE) + High-$\ell$ (SPT+ACT) data \cite{Ade:2013zuv}. Different lines correspond to textures (solid red line), SL strings (dashed black line), and AH strings (dot-dashed blue line). Normalizations correspond to the values given in the leftmost columns of Table~\ref{gmu}.}
\end{figure}

In Fig~\ref{95_Defects} we show spectra obtained using field theory simulations \cite{Bevis:2010gj,Urrestilla:2007sf}. There it can be seen how the constraints derived in the temperature channel affect the possible contribution of each type of defect in B-modes. The most dramatic effect is suffered by AH cosmic strings; their contribution is so suppressed by temperature that it is highly unlikely that they could be a primary source of B-modes. On the other hand, limits imposed on textures and SL strings are not so drastic.

Usually the possible defect contribution is encoded in two related parameters. The first one is $G\mu$, where $G$ is Newton's constant and $\mu$ is the string tension. It encapsulates the most general properties of the defect network, since $\mu$ is strictly linked to the symmetry breaking energy scale\footnote{For textures the string tension parameter does not make much sense. In that case we use $G\mu = \pi\eta^2$, where $\eta$ is the vacuum expectation value, and they are usually parametrized by $\epsilon=4\pi G\eta^2$}. The second is $f_{10}$, is used only in CMB contexts and measures the relative contribution of defects at multipole $\ell = 10$ in the temperature power spectrum. Although there is no direct analytic correspondence between these two parameters, roughly $f_{10} \propto (G\mu)^2$ for $f_{10}\ll 1$. 

\section{Fitting BICEP2 data using defects and inflationary tensor modes}

\begin{figure}[h!]
{\includegraphics[width=0.5\textwidth]{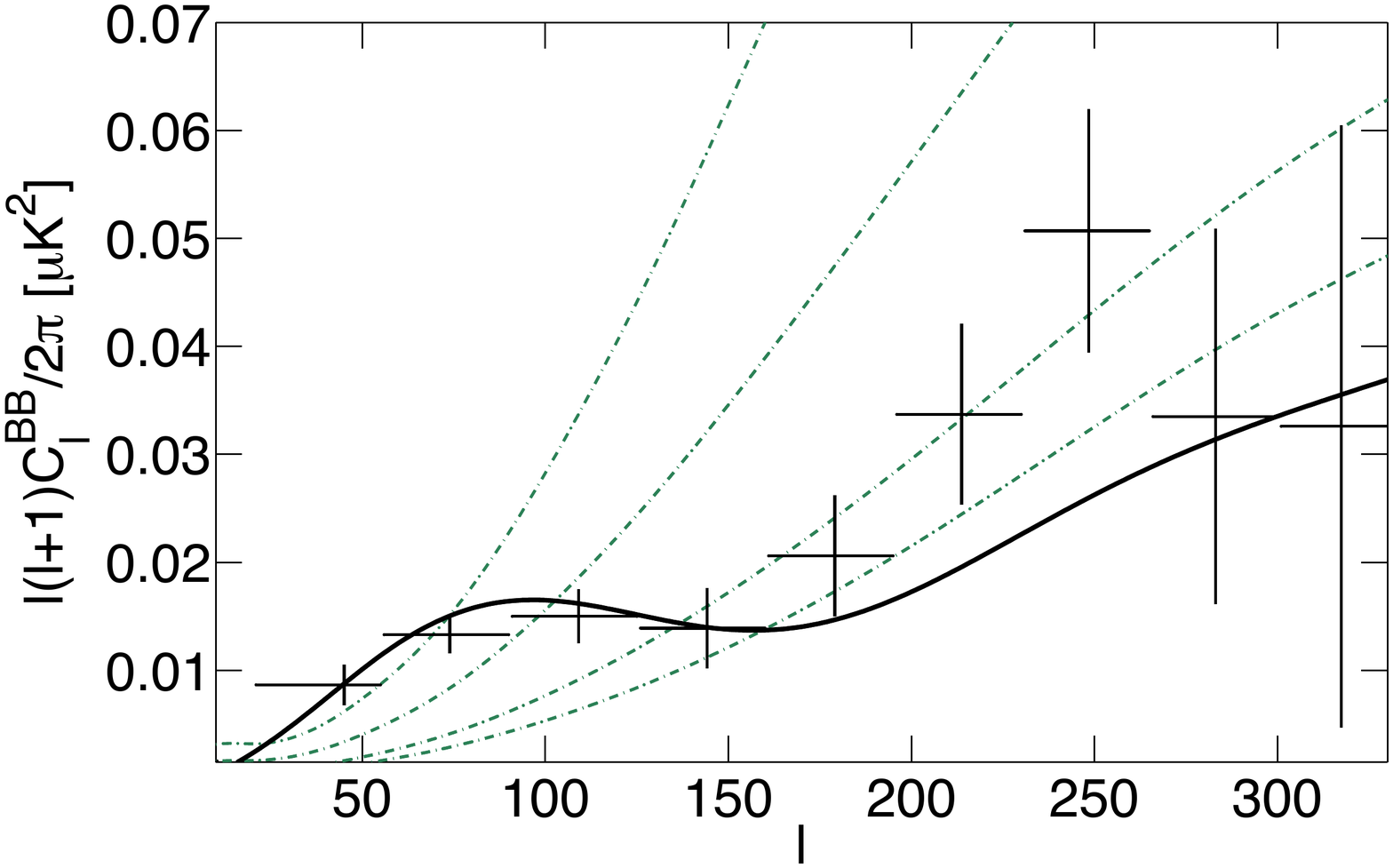}}
{\includegraphics[width=0.5\textwidth]{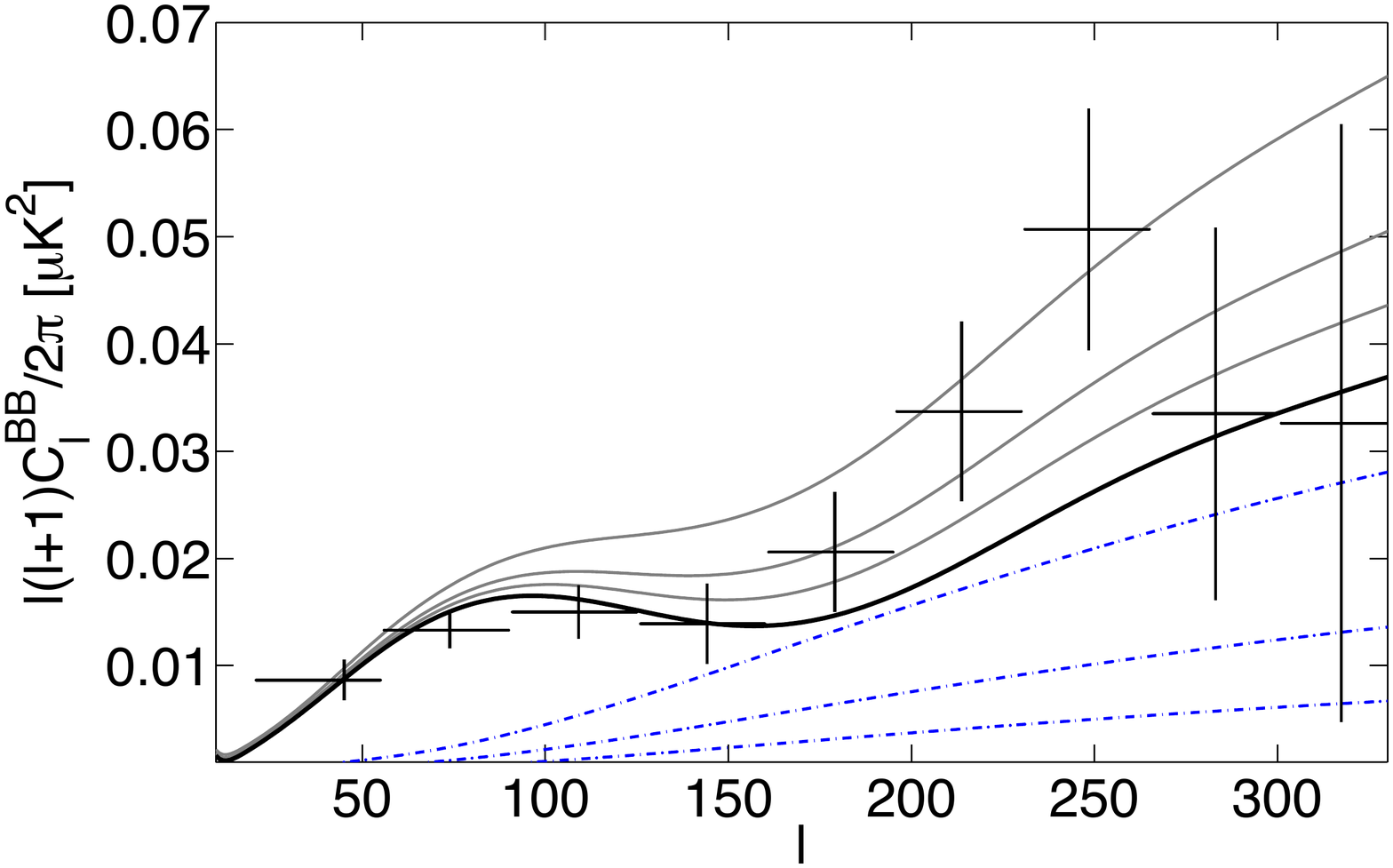}}
\caption{\label{Defects_alone} Comparison of different B-mode spectra and BICEP2 \textit{BB} data. The black curve in both panels represents the best-fit gravitational wave case. In the left panel green-dotted curves show AH string contribution at different normalizations ($f_{10}$ = 0.3, 0.15, 0.06, and 0.03). In the right panel, in turn, we add a contribution from strings (blue dashed) to the inflationary tensor prediction (total spectra in grey). From bottom to top the string fractions are 0.015, 0.03, 0.04 (highlighted in red), and 0.06. All spectra contain lensing contribution.}
\end{figure}

The main question faced by this work is whether defects can account for the whole signal measured by the BICEP2 experiment. In the left panel of Fig.~\ref{Defects_alone} we have included BICEP2 points as well as the best-fit gravitational wave spectrum (solid black line) and the possible contributions of AH cosmic strings considering different normalizations. This figure shows qualitatively that a model containing only defects give a poor fit to the data points. If one tries to fit high-$\ell$ points, the points at low multipoles are underestimated; and vice versa, fitting low multipoles implies overestimating the signal at high-$\ell$s. The analysis of textures and SL strings leads to similar conclusions.

We promoted our basic and qualitative statements into quantitative results via a Monte Carlo analysis. A summary of the analysis can be found in Table~\ref{onlyBICEP}. It can be clearly seen that none of the defects is able to give a comparable fit to the one provided by inflationary tensor modes.

\begin{table*}[t]
\begin{center}
\renewcommand{\arraystretch}{1.2}
\begin{tabular}{|c||c|c|c||c|c|c||c|} 
\hline
Dataset &  \multicolumn{7}{c|}{BICEP2 (only {\it BB})} \\
\hline
Model & \multicolumn{3}{c||}{Defects} &  \multicolumn{3}{c||}{Defects + $r$} & $r$ \\ 
($\Lambda$CDM+) & \multicolumn{3}{c||}{} &  \multicolumn{3}{c||}{} &   \\  \hline
 Param & AH  & SL & TX & AH  & SL & TX& -\\ \hline
$r$ & - & - & - & $0.14_{-0.06}^{+0.04}$ &  $0.14_{-0.06}^{+0.04}$& $0.14_{-0.06}^{+0.04}$   &$0.21_{-0.05}^{+0.04}$\\ 
$10^{12}(G\mu)^2$ &   $0.40_{-0.08}^{+0.07}$& $1.73_{-0.32}^{+0.29}$   & $0.86_{-0.16}^{+0.14}$ & $0.20_{-0.09}^{+0.08}$& $0.87_{-0.39}^{+0.34}$& $0.43_{-0.20}^{+0.17}$  & -\\ 
\hline 
$-\ln{\cal L}_\mathrm{max}$ & $8.1$ & $7.4$ & $6.8$ & $1.6$ & $1.6$ & $1.8$  & $4.3$ \\\hline
 \end{tabular} \\ 
 \caption{\label{onlyBICEP} Parameter estimations and best-fit likelihood values for various cosmological models, fitting for the BICEP2 data. Only the B-mode is used for these estimations.}
\end{center} 
\end{table*} 

Having discarded the role of cosmic defects as primary source of B-modes, at this point we change the perspective and explore how they could assist, as a secondary player, the primary contribution coming from inflationary tensor modes. Defects peak at smaller scales, \textit{i.e.} contribute more significantly at higher multipoles than gravitational waves (see right panel of Fig.~\ref{95_Defects}), hence a mixture of both ingredients could improve the fitting. Results can be found in Table~~\ref{onlyBICEP}, where the maximum likelihoods show how the fit is improved, although it should be noted that this model takes into account two extra contributions.

\section{... and dust}

It has become increasingly apparent that the measurements of BICEP2 might have a non-negligible contribution from polarized dust foregrounds \cite{Mortonson:2014bja,Flauger:2014qra,Adam:2014bub}. Such an astrophysical effect could have been underestimated in the region where BICEP2 made its measurements and would acquire more importance than previously expected.

Following the recipe given in those works and assuming a power-law profile for the dust power spectrum, we extend our previous analysis including dust into the statistical analysis. The dust spectrum has been characterized in the following way:
\begin{equation}
C_\ell^{BB,{\rm dust}} = A_{\rm dust} \ell^{-2.3}
\label{dustPL}
\end{equation}
where $A_{\rm dust}$ is used as the normalization parameter that controls the dust contribution\footnote{Another parametrisation of dust is used in the literature, given by $\Delta^2_{BB}$, which is related to ours via $$\Delta^2_{BB,{\rm dust},\ell}=\frac{\ell^2}{2\pi} C_\ell=\frac{A_{\rm dust}}{2\pi}\ell^{-0.3}$$}$^,$\footnote{While this work was in preparation the Planck collaboration submitted a paper \cite{Adam:2014bub} where they updated the initial dust spectrum to $C_l^{BB}\propto \ell^{-2.45}$. We tested our results and they do not change significantly with this new power-law.}. 

\begin{figure}[h!]
\includegraphics[width=0.6\textwidth]{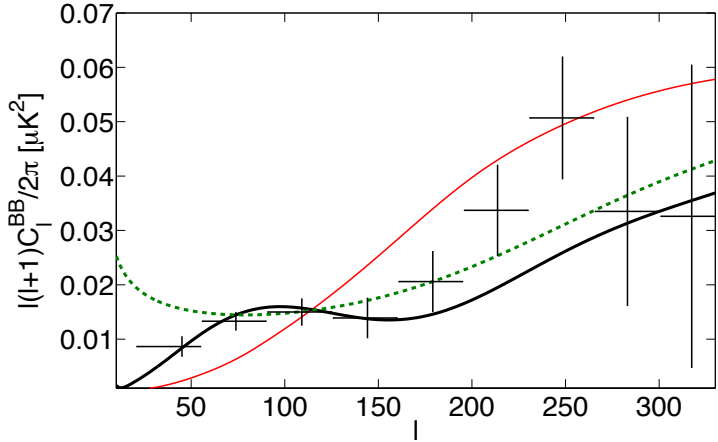}\hspace{2pc}%
\begin{minipage}[b]{13pc}\caption{\label{BestFitDust} BICEP2 \textit{BB} data points and B-mode spectra of tensors (thick solid black line), dust (thick dashed green line) and AH strings (thin solid red line) plus the corresponding lensing contribution. The spectra are obtained using values given in Tables~\ref{onlyBICEP} and ~\ref{onlyBICEPDust}.}
\end{minipage}
\end{figure}


Dust foreground B-mode spectrum, in contrast to the defect case, has more importance at low-$\ell$s (see Fig.~\ref{BestFitDust}). Therefore, dust apparently is more in competition with gravitational waves than with defects. The results are shown in Table~\ref{onlyBICEPDust}. The first thing we notice is that dust alone does a very good job, and even more, it is able to improve the fit given by inflationary tensor modes, confirming what previous works suggested \cite{Mortonson:2014bja,Flauger:2014qra}. 

The two rightmost columns of Table~\ref{onlyBICEPDust} show that a combination of gravitational waves and dust is \textit{not} preferred over a model with dust alone. Moreover, analyzing the best-fit point, one could see that data prefers a model with $r\approx0$. As happened in the previous analysis, defect inclusion improves the fitting, since they assist at higher multipoles.

\begin{table*}[h!]
\begin{center}
\renewcommand{\arraystretch}{1.2}
\begin{tabular}{|c||c|c|c||c||c|} 
\hline
Dataset &  \multicolumn{5}{c|}{BICEP2 (only {\it BB})} \\
\hline
Model & \multicolumn{3}{c||}{Defects + dust} & $r$ + dust &  dust \\ 
($\Lambda$CDM+) & \multicolumn{3}{c||}{} &  &  \\  \hline
 Param & AH & SL & TX & - & -\\ \hline
$r$ & - & - & - &$< 0.22$ & - \\ 
$10^{12}(G\mu)^2$ &   $0.17_{-0.10}^{+0.08}$  & $0.74_{-0.40}^{+0.40}$ & $0.37_{-0.24}^{+0.16}$ & - & -\\ 
$A_{\rm dust} \,[\mu K^2]$ & $0.20_{-0.08}^{+0.06}$ & $0.20_{-0.08}^{+0.06}$ & $0.19_{-0.09}^{+0.06}$ &$0.19_{-0.10}^{+0.10}$  & $0.30_{-0.07}^{+0.06}$ \\
\hline 
$-\ln{\cal L}_\mathrm{max}$ & $1.7$ & $1.7$  & $1.8$ & $3.3$ & $3.3$ \\\hline
 \end{tabular} \\ 
  \caption{\label{onlyBICEPDust} Parameter estimations and best-fit likelihood values for different cosmological models, fitting for the BICEP2 data. This is similar to Table~\ref{onlyBICEP}, but in this case a dust model is included.}
\end{center} 
\end{table*}


\section{New constraints}
We observe that dust is the globally-preferred ingredient by the data \cite{Lizarraga:2014xza}. That is the reason why we derive our latest constraints for defects using a model containing	 defects and dust contribution. In order to obtain the most accurate constraints, we consider a wider CMB dataset, which not only contains B-mode data by BICEP2, but also includes other channels: Planck (TT), WMAP (TE and EE) and high-$\ell$ SPT and ACT (TT).

In Table~\ref{gmu} a comparison of the current constraints and the constraints provided by the Planck collaboration is shown. Constraints of our work have been derived for a $\Lambda$CDM+defects+dust model and using the full-CMB dataset \cite{Lizarraga:2014xza}, whereas the Planck collaboration do it for a $\Lambda$CDM+defects using the same dataset as us except for the BICEP2 \cite{Ade:2013zuv}.
\begin{table}[h!]
\renewcommand{\arraystretch}{1.2}
\begin{center}
\begin{tabular}{|l|c|c||c|c|}
\hline
 & \multicolumn{2}{c||}{Planck col.} & \multicolumn{2}{c|}{This work} \\
\hline
at $<95\%$ C.L. &	 $G\mu$ & $f_{10}$&	 $G\mu$  & $f_{10}$ \\
\hline
Abelian Higgs strings &$3.2\times 10^{-7}$  & 0.024 &$2.7\times 10^{-7}$  & 0.019\\
Semilocal strings & $11 \times 10^{-7}$& 0.041& $9.8 \times 10^{-7}$& 0.031\\
Textures  & $11 \times 10^{-7}$  & 0.054& $7.3 \times 10^{-7}$  & 0.026\\
\hline
\end{tabular}
\caption{\label{gmu} $95\%$ confidence limits for G$\mu$ and $f_{10}$ derived by the Planck collaboration and this work. We use a combined model with dust and defects to obtain current constraints.}
\end{center}
\end{table}

The inclusion of the B-modes measured by BICEP2 does not change drastically the defect constraints, though their inclusion has slightly tighten them.

\section{Conclusions}
Having investigated the possibility of defects being responsible of the recently detected B-mode polarization by the BICEP2 collaboration, we find qualitative and quantitative evidence to assert that defects on their own are a poor fit to the signal. Hence, the need of an additional contribution (gravitational waves or polarized dust) is manifested.  

Nevertheless, defects can help those primary contributions to lift the spectrum at high multipoles. The analysis shows that in both cases, when a defect contribution is added to gravitational waves or dust, the overall fit is improved. 

On the other hand, dust contamination is in general preferred by the data over a gravitational wave source. This is the reason why we choose dust as the primary player in the defect constraint derivation. We calculate new constraints using a full CMB dataset (Planck, WP, High-$\ell$ and BICEP2). New constraints, in Table~\ref{gmu}, are tighter than the ones provided by the Planck collaboration, evidencing the importance of current B-mode experiments in constraining defect models.

\section*{Acknowledgments}
JL and JU acknowledge support from the Basque Government (IT-559-10), the University of the Basque Country UPV/EHU (EHUA 12/11), MINECO (FPA2012-34456) and Consolider Ingenio (CPAN CSD2007-00042 and EPI CSD2010-00064). DD and MK acknowledge financial support from the Swiss National Science Foundation. MH and ARL acknowledge support from the Science and Technology Facilities Council (grant numbers ST/J000477/1, ST/K006606/1, and ST/L000644/1).

\section*{References}
\bibliography{CosmicStrings.bib}

\end{document}